\documentstyle[11pt,aaspp4]{article}
\tighten
\singlespace
\received{19 June 1995}
\accepted{20 July 1995}

\slugcomment{Ap. J. ms. number 32821, accepted}

\lefthead{Kim and Demarque}
\righthead{Theoretical Rossby number}

\begin{document}

\title{The theoretical calculation of the Rossby number\\
       and the `non-local' convective overturn time \\
       for pre-main sequence and early post-main sequence stars}

\author{Yong~-Cheol Kim and Pierre Demarque}

\affil{Department of Astronomy, and Center for Solar and Space Research\\
Yale University, Box 208101, New Haven, CT~06520-8101 \\
E-Mail: kim@astro.yale.edu, demarque@astro.yale.edu}

\begin{abstract}
This paper provides estimates of convective turnover time scales for
Sun-like stars in the pre-main sequence and early post-main sequence phases
of evolution, based on up-to-date physical input for the stellar models.
In this first study, all models have solar abundances, which is typical of
the stars in the Galactic disk where most of the available data have been
collected.  A new feature of these models is the inclusion of rotation in
the evolutionary sequences, thus making it possible to derive theoretically
the Rossby number for each star along its evolutionary track, based on its
calculated rotation rate and its local convective turnover  time near the
base of the convection zone.  Global turnover times are also calculated for
the complete convection zone.  This information should make possible a new
class of observational tests of stellar theory which were previously
impossible with semi-empirical models, particularly in the study of stellar
activity and in research related to angular momentum transfer in stellar
interiors during the course of stellar evolution.
\end{abstract}


\clearpage

\section{Introduction}
There is a rapidly growing body of observations relating to the study of
solar-stellar phenomena which will lead to a better understanding of the
internal dynamics of the pre-main sequence and early post-main sequence
evolutionary phases of stars like the Sun.  But before this wealth of data
can be fully understood, the role of the convection zone structure, its
depth and overturn time scale, and its interaction with rotation need to be
clarified.  This is of importance for understanding not only the mechanism
of angular momentum transfer in stars, but also the evolution of rotating
stars, both in the pre main-sequence and post main-sequence phases. The
interaction of rotation with convection is now widely believed to be
responsible for the generation of stellar dynamos and the observed stellar
magnetic activity and activity cycles.

Durney and Latour (1978) (see also Durney, Mihalas and Robinson 1981, and
Durney and Robinson 1982) made an important step forward in relating the
principles of mean-field dynamo theory to the observations. They showed
that if a stellar dynamo is responsible for the observed stellar activity,
there should be a relation between magnetic activity and the
characteristics of rotating stellar convection zones.  With the help of
dimensional arguments, they pointed out the significance of the Rossby
number (proportional to the ratio of the rotation period to the convective
turnover time) in the dynamo mechanism.  Since in the dynamo model, the
dynamo action is believed to take place at the base of the convection zone,
anchored in the radiative layers just below the convective interface, the
convective turnover time of the deepest part of the convection zone is the
most relevant in the evaluation of the Rossby number.  Soon afterwards, the
availability of precise rotation periods for magnetically active stars
(Baliunas et al. 1983) made it possible to test this hypothesis in a
semi-empirical way by combining convective turnover times derived from
model convection zones, such as the models of Gilman (1980), with the
observed rotation periods (Noyes 1983).  This was done by a number of
researchers (Mangeney and Praderie 1984;  Hartmann et al. 1984; Noyes et
al. 1984) for stars near the main sequence and for pre-main sequence stars
(Simon et al. 1985).  Since then a large number of studies have been
performed correlating different types of magnetic activity indices to such
semi-empirically derived Rossby numbers and to other parameters such as the
stellar rotation  rate (Basri 1987; Shrijver and Rutten 1987; Simon and
Fekel 1987; Dobson and Radick 1989).

The pattern of convective velocities as a function of effective temperature
and age derived from  stellar models, which serve  as input in Rossby
number calculations, depend sensitively on the particulars of the stellar
interior models, either input parameters such as chemical composition, or
mass, or physics input such as opacities or the equation of state used.  In
addition, because of the well-known non-linearity of the equation of
stellar structure, it is unadvisable to construct stellar envelope models
by simple inward integration without applying the interior boundary
conditions, as the early calculations  frequently did; fully consistent
interior models are needed.  An important step in relating activity
observations to self consistent stellar evolutionary tracks and the
predicted evolutionary changes in convective overturn times was made by
Gilliland (1985, 1986).  Other calculations of Rossby numbers,  based on
complete  main sequence stellar models, have also been published by
Rucinski and VandenBerg (1986, 1990).  Since then, rapid progress has been
made in our knowledge of stellar opacities and equation of state (see e.g.
Rogers and Iglesias 1994), and much improved models of the Sun and Sun-like
stars can be constructed (Guenther et al 1992; Chaboyer et al. 1995;
Guenther, Kim and Demarque 1995).

The theory of rotating stellar evolution has also advanced.  The work of
Endal and Sofia (1978, 1981), which included the spin-down due to a stellar
wind, and introduced into stellar evolution the effects of various
rotationally induced mixing processes acting on different time scales, thus
relating angular moment transfer to internal mixing, opened up new ways of
confronting theory and observation (Pinsonneault et al. 1989, 1990;
Chaboyer et al. 1995).

The purpose of this paper is to provide estimates of turnover time scales
for Sun-like stars in the pre-main-sequence and early post main-sequence
phases of evolution, based on up-to-date physical input for the stellar
models.  In this first study, all models have solar abundances, which are
typical stars in the Galactic disk. Another new feature of these models is
the inclusion of rotation.  Because the evolution of internal rotation has
been included in the models, it is possible to derive theoretically the
Rossby number for each star along its evolutionary track based on the
theoretical estimates for both the convective turnover time and the
rotation rate of the convection zone.  These internally self consistent
models  should make possible a new class of observational tests of stellar
theory which were impossible with semi-empirical models.

We describe convection by the mixing-length formalism in the usual way.
While it is known that the convective velocities near the stellar surface
are not well described by the mixing-length approximation (Kim et al.
1995a,b), the convective turnover time scales calculated here are dominated
by the conditions near the base of the convection zone, where the
temperature gradient is for all practical purposes adiabatic, and the
mixing-length approximation is known to provide an adequate
description of convection, at least in an average sense (Chan and Sofia
1989; Lydon et al. 1992).  For this reason, the Rossby number estimates
should  be little affected (subject to a constant scale factor) by
improvements in our understanding of convection. We emphasize, however,
that for many other purposes, such as describing the outer layers where
radiation plays a dominant role, the mixing length approximation is
inadequate, and more refined convection models that take into consideration
the interplay between convection and radiation, are needed (Kim et al.
1995a,b).  This conclusion applies in particular for understanding the
structure of the transition superadiabatic layer at the top of the
convection zone.  It is also likely to apply for understanding the behavior
of magnetic fields, the details of the generation of acoustic noise in
stellar chromospheres, and the driving of p-modes in Sun-like stars.

Section 2 describes the series of stellar models with masses ranging from
$0.5 M_\odot$ to $1.2 M_\odot$ which were evolved from the fully convective
pre-main-sequence Hayashi phase to the sub-giant phase. The calculation of
the convective turnover time and of the Rossby number and their evolution
as a function of time are considered in Sections 3 and 4, respectively.
Finally, we briefly discuss the results  in section 5.

\section{Calculations}
\subsection{Stellar models}

A series of stellar models with masses ranging from 0.5 to 1.2$M_\odot$  (in
0.1$M_\odot$ increments), have been evolved from a fully convective
pre-main sequence model to the sub-giant phase.  The OPAL opacities tables
(Iglesias and Rogers 1991), constructed for the solar mixture of Anders and
Grevesse(1989) were used, together with the Kurucz (1991) low temperature
opacities.  The Kurucz (1992) model atmospheres served  as surface boundary
conditions.  The numerical tolerances and input physics were identical for
all evolutionary runs, and similar to those adopted by Chaboyer et al.
(1995). All models used the parameters derived for the standard solar
model,  where the initial $X$, $Z$, and the mixing length ratio $\alpha$
are varied until a solar model at the solar age of 4.55Gyr (Guenther 1989)
has the observed solar values of luminosity, radius, and $Z/X$.  In
addition, the solar surface rotational velocity and ${}^7Li$ depletion were
used to calibrate the rotation and diffusion parameters of all evolutionary
sequences, as described in sections 2.2 and 2.3 below.  The solar model in
this calibration matches the solar radius and luminosity to within 0.01\%,
while the surface $Z/X$ matches the observed value to within 1.0\%.  The
model also reproduced the observed solar rotation rate and $Li$ depletion
to within 1.5\%.  Table 1 summarizes the characteristics of the models and
their input parameters.

Figure 1 shows the evolutionary tracks in the H-R diagram. For the
internal rotation rates considered here, rotation has a negligible
effect on both the rate of evolution and the path of the evolutionary
track in the H-R diagram (Pinsonneault et al. 1989; Deliyannis et al. 1989).

\subsection{Rotation}

 All models used in this paper have been constructed using a version of the
Yale Rotating Stellar Evolution Code (Prather 1976, Pinsonneault 1988).
Recently, the YREC has been improved in the microscopic diffusion and its
interaction with rotational mixing (Chaboyer et al. 1995).   The
calculation has been carried out using this improved version.

The evolutionary sequences were started from fully convective pre-main
sequence models in the Hayashi phase.  At first, the whole  star rotates as
a rigid body, and spins up as it contracts.  The torque due to the
stellar wind then takes over and spins the star down (see below),
and in the process of transferring out internal angular momentum,
progressively depletes ${}^7Li$ in the star.
The evolution of the internal angular momentum distribution follows the
approach of Pinsonneault et al (1989), which includes the effects of
rotationally induced instabilities in the radiative layers.
This results in a state of differential rotation.  The Sun seems to be
overdepleted in ${}^7Li$ compared with other stars of its age and spectral
class.  Since the amount of mixing in a star increases as its initial
rotational velocity increases, it is likely that the Sun was initially a
rapid rotator.  Therefore, even though observations show that the majority
of T-Tauri stars have rotation velocities around $10 km/s$,  we have chosen
an initial rotation velocity of $30 km/s$ for the solar model.
Observations of T-Tauri stars also indicate that there is no large
difference in surface rotation rates between high and low mass stars over
the mass range we study, i.e. $0.5 \sim 1.2M_\odot$ (Bouvier 1991; Bouvier
et al. 1993), and for this reason, we have the applied the same rotation
parameters to all masses.  Figure 2 shows the evolution of the rotation
period of our theoretical models. It is important to note that changing the
initial rotation velocity affects the time-scale for early spin-down, but
does not change appreciably the final configuration.  This is due to the
fact that our calibration requires the solar model to rotate at the present
rotation rate of the Sun (the adopted initial rotational velocity affects
primarily only the present ${}^7Li$ abundance, which is not particularly
relevant to the convective turnover calculations of this paper).  Note in
Figure 2 that the rotation period depends sensitively on mass for a given age.

When calculating the evolution of the Rossby number, the adopted wind law
is the most important input of our rotating models, as internal structural
effects of rotation are minimal in these models which rotate relatively
slowly.  A modified version of Kawaler's parameterization for the loss of
angular momentum due to magnetic stellar wind (Kawaler 1988) has been used
(Chaboyer et al. 1995).  It is given by:
$$
{ {dJ} \over {dt} }= f_k K_w
{R}^{2-N}
{M}^{-N/3}
{\dot{M}}^{1-2N/3}
\omega^{1+4N/3} \qquad (\omega < \omega_{crit}),
$$
$$
{ {dJ} \over {dt} }= f_k K_w
{R}^{2-N}
{M}^{-N/3}
{\dot{M}}^{1-2N/3}
\omega \omega_{crit}^{4N/3} \qquad (\omega \geq \omega_{crit}),
$$
where R is the radius in units of the solar radius ($R_\odot$),
M is the mass in units of the solar mass ($M_\odot$),
$\omega_{crit}$ introduces a saturation level into the angular
momentum loss law (set to $1.5 \times 10^{-5} \ s^{-1}$),
$K_w = 2.036 \times 10^{33} \left(1.452 \times 10^{9}\right)^N$
in cgs units, and $\dot{M}$ is the mass-loss rate in unit of
$10^{-14} M_{\odot} yr^{-1}$ (set to $2.0$).
Here, it is primarily the exponent $N$ in the wind model (a measure of the
magnetic field geometry), which determines the rate of angular momentum
loss with time.  We have adopted the value of 1.5 for $N$, which reduces
to the empirical Skumanich (1972) law near the main sequence i.e.,
$$ v_{rot} \propto \tau^{-.51},$$
where $ v_{rot}$ is equatorial rotation velocity, and $\tau$ is the age
in Gyr.
The constant factor in the wind model, $f_k$, determines the total amount
of the angular momentum loss. We adjust $f_k$ for a given $N$ to give the
solar surface rotation velocity at the solar age. We use the observed value
at about 30 degrees from the equator, $1.86 km/s$, since this value is
close to the mean value which from the seismology data,  the interior of
the Sun appears to approach (Libbrecht and Morrow 1991)

For the sake of completeness, the transport of angular momentum and
chemical elements have also been taken into account in the models.   Two
types of   rotation-induced mixing -- the dynamical shear instabilities and
the Solberg-Hoiland instability -- and three type of secular instabilities
-- meridional circulation, the Goldreich-Schubert-Fricke instability, and
the secular shear instability -- were included in the calculations
(Chaboyer et al. 1995, Pinsonneault et al. 1989).

In this study, the uncertain effects of other secular instabilities are
treated as free parameter, and the diffusion coefficients are set to be the
same as that of meridional circulation, $f_c$.   The value of $f_c$ is
fixed in the solar model by requiring the model lithium depletion to match
the solar value at the solar age. The depletion of $Li$ is inferred by
comparing the cosmic abundance with the photospheric abundance.  A
comparison of the meteoric abundance with the photospheric abundance shows
the depletion to be a factor of $140^{+40}_{-30}$ (Anders and Grevesse
1989).  We have therefore set the solar ${}^7Li$ depletion factor equal to
140.

\subsection{Diffusion}
The microscopic diffusion coefficients of Michaud and Proffitt (1993) have
been used.  They have the advantage of being valid not only for ${}^4He$
and ${}^1H$, but also  for ${}^3He$, ${}^6Li$, ${}^7Li$, and ${}^9Be$.
Comparison with the diffusion coefficients of Thoul et al. (1994) indicates
that the Michaud-Proffitt coefficients are good to within 15\%. Finally we
note that when diffusion is taken into account, the surface $Z/X$ is not a
constant during a stellar evolution calculation. As ${}^4He$ diffuse with
respect to hydrogen with the relatively short time scale, the model
structures are affected. Measurements of the solar photosphere do not
actually determine $Z$ -- they measure $Z/X$ (i.e. $[Fe/H]$).  The Anders
and Grevesse (1989) photospheric mixture with meteoritic Fe gives
$Z/X=0.0267\pm0.001$.  Thus, our model of the present Sun was constrained
to match this number.

\section{Global turnover time and Rossby number}

 The close connection between stellar rotation and its chromospheric
emission can be described in terms of general stellar dynamo models. In the
mean-field dynamo theory, a dimensionless parameter, the dynamo number,
characterizes the model behavior. The dynamo number is essentially
proportional to the inverse square of the Rossby number, $N_R$, which is
the ratio of the stellar rotation period to the local convective turnover
time (Durney and Latour 1978; Noyes et al 1984).
Thus, in principle, one could draw a theoretical Rossby number vs.
magnetic activity diagram.

In practice, however, our knowledge on stellar convection is too limited to
calculate `correct' convective turnover times. The characteristic length
scales as well as the velocities are not well known. Even when one decides
to resort to the mixing length approximation, there are still
uncertainties:  the  mixing length ratio $\alpha$  is assumed to be the
same for all stars with different masses and/or at different evolutionary
stage, which is probably not be quite correct.  In addition, some
assumption must be made as to where in the convection zone the dynamo
process is operating, since the convective overturn time is strongly depth
dependent. For example, Gilman (1980) set the characteristic convective
overturn time equal to the convective overturn time one scale height above
the bottom of the convective zone. On the other hand, Gilliland (1986)
determined the turnover time at a distance of half of the mixing length
above the base of the convection zone.

\subsection{Global convective turnover time}
To depict the characteristics of convection at each stellar evolution
stage, two parameters have been calculated; the `global' convective
turnover time, and the Rossby number. For the characteristic time scale of
convective overturn, the `non-local' (or `global') convective turnover time
has been calculated  at each time step.  It is:
$$ \tau_c = {\int^{R_\ast}_{R_b}} {{dr} \over {v}} $$
where ${R_b}$ is the location of of the bottom of the surface convection
zone, which is defined where the ${\nabla - \nabla_{ad}} = 0$, ${R_\ast}$
is the total radius of the stellar model, and $v$ is the local convective
velocity.   Figure 3 shows the evolution with time of the convective
turnover time.  Note that in the pre-main sequence phase, $\tau_c$ varies
rapidly with time.  Near the main sequence, $\tau_c$ remains nearly
constant and is primarily a function of mass.

\subsection{Rossby number}
For the Rossby number calculation,
the characteristic convective overturn time was set equal to the `local'
convective overturn time at a distance of half of the mixing length
${{\alpha \  H_P} \over {2} }$ above the base of the convection zone.
The ratio of the rotation period to this characteristic convective
overturn time is used to characterize the Rossby number in the deep
convection zone, where dynamo generation of magnetic fields is thought
to occur.
$$N_R = 2 \pi v / \alpha \Omega H_P$$
where $v$ is the characteristic convective velocity, $\alpha$ is the
mixing length ratio, $\Omega$ is rotational velocity, and
$H_P$ is the local pressure scale height.   It turns out that {\bf the
evolution of the `local' turnover time is the same as the `non-local'
one, except for a scaling factor}.
The evolution of the inverse squared Rossby number is illustrated in
Figure 4.  This quantity, sometimes called the `dynamo number', is
believed to be proportional to the strength of magnetic activity.
Note that the dynamo number depends on both the age and the mass of
the star.

\section{Isochrones}
Theoretical isochrones offer the opportunity to test stellar evolution
theory in star clusters where stars are coeval and formed from a gas
cloud of uniform composition.  Conversely, when properly calibrated,
isochrones can become a powerful tool to study the properties of field stars.

Isochrones were constructed using the evolutionary tracks for the ages of
0.2, 0.5, 0.7, 1.0, 4.55 (the solar age), 10, and 15 Gyr.  Their
characteristics are listed in Table 2.  Figure 5 shows a plot of
isochrones of the non-local turnover time vs. $\log T_{eff}$ (the solid
lines).  For comparison, a few isochrones of the local turnover time are
shown (the dotted lines), in the same figure. Isochrones
of the non-local turnover time vs. rotation period are given in Figure 6.
In Figure 7, rotation period vs. $\log T_{eff}$ is shown, where the
increase of $\log T_{eff}$ can be understood as the increase of the stellar
mass, because of the proximity of the main sequence. The right most point
of each line is for 0.5 $M_\odot$. Assuming our treatment of stellar
rotation is correct, then one can use Figure 7 to uniquely determine
stellar mass and age from the effective temperature and the rotation
period. Figure 8 shows the the inverse square of the Rossby number,
$N_R^{-2}$ vs. $\log T_{eff}$. Once empirical relations between $N_R^{-2}$
and magnetic activity indices are determined, one can use Figure 8 for
determination of the age and the mass  of a star by observing its effective
temperature and an activity index.  Figure 9 is the plot of $N_R^{-2}$ vs.
rotation period, where for an isochrone, the right most point of the line
represent the lowest mass. We see that, given the assumptions implicit in
our discussion, our grid of theoretical evolutionary tracks provide the
means to determine the age and the mass of a star from a measurement of its
rotation period and an activity index.

\section{Discussion}
Estimates of turnover time scales and the Rossby number are provided, for
Sun-like stars in the pre-main sequence and early post-main sequence phases
of evolution, based on up-to-date physical input for the stellar models,
and including rotation.

We expect the results in this paper to be robust, since the convective
turnover timescale is weighted toward the deepest part of the convection
zone, where the shortcomings of the mixing length approximation are least
important. This is the reason why our `global' and `local' convective time
scale give the same result except for a scaling factor
(e.g. Figure 5).
This is consistent with recent numerical simulations of
convection (e.g. Chan and Sofia 1989;  Kim et al. 1995b) which confirm the
validity of the mixing length approximation  in the limit of deep and
efficient convection.

In this study, all models have solar abundances; they will therefore find
applications in the interpretation of the rotational history and magnetic
activity indices for stars in young star clusters and Sun-like field stars,
which are the most common stars in our part of the Galactic disk.  Caution
must be exercised, however, with stars with chemical composition that
differ appreciably from solar.  Both the depth of the convection zone and
the convective velocities are known to depend on opacities and equation of
state.  As more detailed observations about the rotational properties and
magnetic activity of very metal-poor and very metal rich stars become
available, the sensitivity of convective turnover timescales to chemical
composition parameters will need to be explored in some detail.

\acknowledgments
YCK would like to thank B. Chaboyer and M. Pinsonneault.  This work was
supported in part by NASA grants NAG5-1486 and NAGW-2469 to Yale
University.

\clearpage

\begin{center}
{\bf Figure Captions}
\end{center}
\begin{description}
\item[Figure 1:]  The evolutionary tracks in the theoretical HR-diagram.
Several isochrones have also been drawn for 1 Gyr to 15 Gyr.  The
lower age isochrones are indisguishable from the 1 Gyr isochrone on
the scale of this diagram.
\item[Figure 2:]  The rotation period as a function of age and mass.  The
slope near the main sequence of each mass curve corresponds to the
Skumanich law with a proportionality factor which depends on mass.
\item[Figure 3:]  The non-local convective turnover time as a function of age
and mass.  Near the main sequence, $\tau_c$ remains nearly constant
with time for a given mass.
\item[Figure 4:]  The dynamo number ($\propto N_R^{-2}$),
 often used as a measure of magnetic
activity strength, as a function of age and mass.  This plot (together
with Figures 8 and 9)
provides the means of calibrating magnetic activity
indices along the main sequence of a star cluster.
\item[Figure 5:]  The global and local convective turnover times as a
function of effective temperature and age.
\item[Figure 6:]  Global convective turnover time as a function of rotation
period and age.
\item[Figure 7:]  Rotation period as a function of effective temperature and
age.
\item[Figure 8:]  The dynamo number as a function of effective temperature
and age.
\item[Figure 9:]  The dynamo number as a function of rotation period and age.
\end{description}

\clearpage

\begin{deluxetable}{ll}
\tablewidth{0pt}
\tablecaption{Input physics}
\tablehead{ \colhead{Parameter} &  \colhead{Input} }
\startdata
Mass & $0.5 \sim 1.2{{M}_{\odot}}$ \nl
Mixing length ratio  & 1.86315 \nl
Weight fraction of hydrogen, X  & 0.70952 \nl
Weight fraction of all heavy elements, Z  & 0.01926 \nl
Mixture of heavy elements  & Anders-Grevesse(1989) \nl
The exponent in the wind model, $N$ & 1.5 \nl
The constant factor in the wind model, $f_k$ & 17.4837 \nl
The diffusion coefficient, $f_c$ &0.05575\nl
Initial rotation velocity & $30 \ km\ s^{-1}$ \nl
Opacity tables & OPAL \nl
 & with Kurucz opacity tables for low temperature \nl
Atmosphere & Kurucz model atmosphere \nl
Equation of state &  The standard implementation \nl
 & with Debye-H\"uckel correction \nl
The microscopic diffusion & Michaud and Proffitt (1993)
\enddata
\end{deluxetable}

\clearpage

\begin{deluxetable}{rrrrrrr}
\footnotesize
%\tablewidth{0pt}
\tablewidth{370.0pt}
\tablecaption{Isochrones}
\tablehead{
\colhead{$M/M_\odot$} &
\colhead{$\log T_{eff}$} &
\colhead{$(B-V)$ \tablenotemark{a}} &
\colhead{$\log L/L_\odot$} &
\colhead{$\tau_c$ \tablenotemark{b}} &
\colhead{$N_R^{-2}$ \tablenotemark{c}} &
\colhead{Rotation Period}\nl
 & & & &\colhead{(day)}& &\colhead{(day)}
}
\startdata
\multicolumn{7}{c}{0.20 Gyr}\nl
 0.5& 3.54& 1.55&-1.50&  125.10&    6.18&   23.41 \nl
 0.6& 3.59& 1.37&-1.19&  104.06&   27.58&    9.18 \nl
 0.7& 3.64& 1.16&-0.88&   83.19&   28.40&    7.23 \nl
 0.8& 3.68& 0.99&-0.61&   66.50&   24.13&    6.31 \nl
 0.9& 3.72& 0.83&-0.36&   53.15&   18.90&    5.69 \nl
 1.0& 3.75& 0.69&-0.14&   40.27&   13.30&    5.19 \nl
 1.1& 3.78& 0.57& 0.06&   27.49&    7.73&    4.67 \nl
 1.2& 3.80& 0.49& 0.24&   13.95&    2.48&    4.13 \nl
\cline{1-7}
{}~\\[-8pt]
\multicolumn{7}{c}{0.50 Gyr}\nl
 0.5& 3.53& 1.60&-1.53&  137.40&   11.13&   19.19 \nl
 0.6& 3.59& 1.37&-1.18&  106.56&   13.57&   13.44 \nl
 0.7& 3.64& 1.16&-0.87&   84.84&   11.77&   11.46 \nl
 0.8& 3.68& 0.99&-0.60&   67.18&    9.31&   10.25 \nl
 0.9& 3.72& 0.83&-0.35&   53.07&    7.01&    9.33 \nl
 1.0& 3.75& 0.69&-0.13&   39.97&    4.82&    8.57 \nl
 1.1& 3.78& 0.57& 0.07&   26.97&    2.56&    7.94 \nl
 1.2& 3.80& 0.49& 0.26&   13.14&    0.69&    7.42 \nl
\cline{1-7}
{}~\\[-8pt]
\multicolumn{7}{c}{0.70 Gyr}\nl
 0.5& 3.53& 1.60&-1.53&  140.32&    9.26&   21.32 \nl
 0.6& 3.59& 1.37&-1.18&  107.55&   10.14&   15.69 \nl
 0.7& 3.64& 1.16&-0.87&   84.88&    8.42&   13.59 \nl
 0.8& 3.68& 0.99&-0.60&   67.40&    6.57&   12.22 \nl
 0.9& 3.72& 0.83&-0.35&   52.96&    4.92&   11.12 \nl
 1.0& 3.75& 0.69&-0.13&   40.00&    3.37&   10.21 \nl
 1.1& 3.78& 0.57& 0.08&   26.76&    1.77&    9.44 \nl
 1.2& 3.80& 0.49& 0.27&   12.87&    0.46&    8.81 \nl
\cline{1-7}
\tablebreak
{}~\\[-8pt]
\multicolumn{7}{c}{1.00 Gyr}\nl
 0.5& 3.53& 1.60&-1.53&  141.14&    7.15&   24.45 \nl
 0.6& 3.59& 1.37&-1.18&  108.04&    7.30&   18.61 \nl
 0.7& 3.64& 1.16&-0.87&   85.43&    5.90&   16.29 \nl
 0.8& 3.69& 0.95&-0.59&   67.49&    4.55&   14.68 \nl
 0.9& 3.72& 0.83&-0.34&   52.74&    3.40&   13.37 \nl
 1.0& 3.75& 0.69&-0.12&   39.86&    2.31&   12.28 \nl
 1.1& 3.78& 0.57& 0.09&   26.32&    1.19&   11.34 \nl
 1.2& 3.80& 0.49& 0.29&   12.40&    0.30&   10.56 \nl
\cline{1-7}
{}~\\[-8pt]
\multicolumn{7}{c}{2.00 Gyr}\nl
 0.5& 3.53& 1.60&-1.53&  142.19&    4.11&   32.61 \nl
 0.6& 3.59& 1.37&-1.18&  108.88&    3.73&   26.23 \nl
 0.7& 3.64& 1.16&-0.86&   85.26&    2.91&   23.24 \nl
 0.8& 3.69& 0.95&-0.58&   67.28&    2.22&   20.98 \nl
 0.9& 3.73& 0.78&-0.32&   52.62&    1.63&   19.19 \nl
 1.0& 3.76& 0.65&-0.09&   39.25&    1.07&   17.72 \nl
 1.1& 3.78& 0.57& 0.14&   25.00&    0.53&   16.23 \nl
 1.2& 3.80& 0.49& 0.34&   11.44&    0.13&   15.07 \nl
\cline{1-7}
{}~\\[-8pt]
\multicolumn{7}{c}{4.55 Gyr}\nl
 0.5& 3.53& 1.60&-1.52&  144.87&    1.80&   49.92 \nl
 0.6& 3.59& 1.37&-1.16&  110.05&    1.54&   40.97 \nl
 0.7& 3.64& 1.16&-0.84&   85.10&    1.19&   36.26 \nl
 0.8& 3.69& 0.95&-0.54&   66.70&    0.89&   32.77 \nl
 0.9& 3.73& 0.78&-0.26&   51.21&    0.65&   29.83 \nl
 1.0& 3.76& 0.65& 0.00&   37.49&    0.41&   27.66 \nl
 1.1& 3.78& 0.57& 0.25&   24.82&    0.22&   25.14 \nl
 1.2& 3.78& 0.57& 0.44&   26.42&    0.28&   23.40 \nl
\cline{1-7}
\tablebreak
{}~\\[-8pt]
\multicolumn{7}{c}{10.00 Gyr}\nl
 0.5& 3.53& 1.60&-1.50&  145.61&    0.73&   78.38 \nl
 0.6& 3.59& 1.37&-1.13&  109.32&    0.62&   64.26 \nl
 0.7& 3.65& 1.11&-0.78&   83.95&    0.47&   56.79 \nl
 0.8& 3.70& 0.91&-0.45&   65.26&    0.33&   52.86 \nl
 0.9& 3.74& 0.74&-0.11&   50.58&    0.24&   48.69 \nl
 1.0& 3.75& 0.69& 0.28&   50.94&    0.32&   42.01 \nl
\cline{1-7}
{}~\\[-8pt]
\multicolumn{7}{c}{15.00 Gyr}\nl
 0.5& 3.54& 1.55&-1.48&  144.95&    0.39&  107.04 \nl
 0.6& 3.60& 1.33&-1.10&  108.32&    0.33&   87.34 \nl
 0.7& 3.66& 1.07&-0.73&   83.08&    0.24&   78.18 \nl
 0.8& 3.71& 0.87&-0.34&   64.97&    0.18&   70.55 \nl
 0.9& 3.74& 0.74& 0.13&   68.27&    0.34&   53.99
\enddata
\tablenotetext{a}{ Revised Yale
Isochrones and Luminosity Functions (Green et al. 1987)}
\tablenotetext{b}{ $ \tau_c = {\int^{R_\ast}_{R_b}} {{dr} \over {v(r)}} $,
where ${R_b}$ is the location of the bottom of the surface convection zone,
${R_\ast}$ is the total radius of the stellar model, and $v(r)$ is the
convective velocity as a function of radius.  }
\tablenotetext{c}{ $N_R = 2 \pi v / \alpha \Omega H_P$
where $v$ is the local convective velocity at  a distance of the half of
the mixing length ${{\alpha \  H_P} \over {2} }$ above the base of
the convection zone, $\alpha$ is the
mixing length ratio, $\Omega$ is rotational velocity, and
$H_P$ is the local pressure scale height.}
\end{deluxetable}

\end{document}